\documentclass[12pt]{iopart}
\usepackage{graphicx}
\usepackage{cite}
\pdfoutput=1 % if your are submitting a pdflatex (i.e. if you have
\begin{document}

\title{Tagging $b$ quarks without tracks
using an Artificial Neural Network algorithm}
%\title{\boldmath A title with some math: $x=1$}

%% %simple case: 2 authors, same institution
\author{B. Todd Huffman, Thomas Russell, and Jeff Tseng}
%JHEP control sequence
%\affiliation{Oxford University dept. of Particle Physics,\\Oxford, United Kingdom}
\address{Particle Physics, Oxford University \\ 
	Keble Road \\ Oxford  OX1 3RH \\
	United Kingdom}
\ead{todd.huffman@physics.ox.ac.uk}
\vspace{10pt}
\begin{indented}
\item[] 15 June 2017
%\item[] \today
\end{indented}

\begin{abstract}
Pixel detectors currently in use by high energy physics
experiments such as ATLAS and CMS, are critical 
systems for tagging $B$ hadrons within particle jets. 
However, the performance of standard tagging algorithms 
begins to fall in the case of highly boosted $B$ hadrons 
($\gamma \beta = p/m >200$).
This paper builds on the work of our
previous study that uses the jump in hit multiplicity among the pixel
layers when a $B$ hadron decays within the detector volume.
First, multiple $pp$ interactions within a finite luminous region
were found to have little effect.  
Second, the study has been extended to use the multivariate techniques of an
artificial neural network (ANN). After training, the ANN shows
significant improvements to the ability to reject light-quark and charm jets;  
thus increasing the expected power of the technique.
\end{abstract}

\section{Introduction}
\label{sec:intro}

Many of the most exciting searches for new physics beyond the Standard Model,
as well as further studies of the Standard Model itself, benefit from being
able to identify high-energy jets containing $b$ quarks (``$b$ jets'').
Examples include Higgs pair production and decay via
$HH\rightarrow b\overline{b}b\overline{b}$, sensitive to Higgs trilinear
couplings~\cite{Bishara:2016kjn, Behr:2015oqq}; graviton and radion decays to 
heavy fermions and bosons in warped extra dimension 
models~\cite{Gouzevitch:2013qca}; third-generation superpartners in 
supersymmetry~\cite{Alwall:2008ag}; and indeed any new physics with 
preferential couplings to heavy Standard Model particles or third-generation 
fermions in particular.

Because $B$~hadron tagging is so important, progress has been made by the 
CMS and ATLAS experiments in improving the effiency of their conventional taggers. 
They have been mainly using boosted decision tree techniques and also some artificial 
neural nets to make gains in discriminating power at higher jet energies. The interested 
reader can find this information in references~\cite{CMSbtag}, \cite{ATLASbtag}, 
\cite{ATLASbtag2} (and references contained within these reports). 
In particular we see in figure~25(a) in~\cite{CMSbtag} that the CMS collaboration has 
managed to obtain$~\sim 20 \%$ efficiency at jet $p_T \simeq 800$~GeV while 
figure~5 within~\cite{ATLASbtag} shows better than $60\%$ efficiency but only to 
jet $p_T \simeq 400$~GeV. Even with these improved results it is still clear that maintaining 
high $b$-tagging efficiency out to TeV-scale jet $p_T$ is a difficult problem. 

This work follows from a previous study by two of the current 
authors~\cite{Huffman:2016wjk}.
Section~\ref{sec:rehash} will summarise that previous study. 
Section~\ref{sec:improv} will explain the simulations used, the modifications,
and the improvements to those simulations implemented 
by the authors. 
Section~\ref{sec:Perform} will introduce and explain the implementation of an
Artifical Neural Network (ANN) in order to investigate the level of improvement 
to $b$-tagging efficiency and light quark rejection in multi-TeV jets. We will
show the results of these new studies in section~\ref{sec:ANN} while 
section~\ref{sec:conclusion} concludes.

\section{The ``multiplicity jump'' tagger}
\label{sec:rehash}

\begin{figure}[thbp]
\centering 
\includegraphics[height=.6\textwidth]{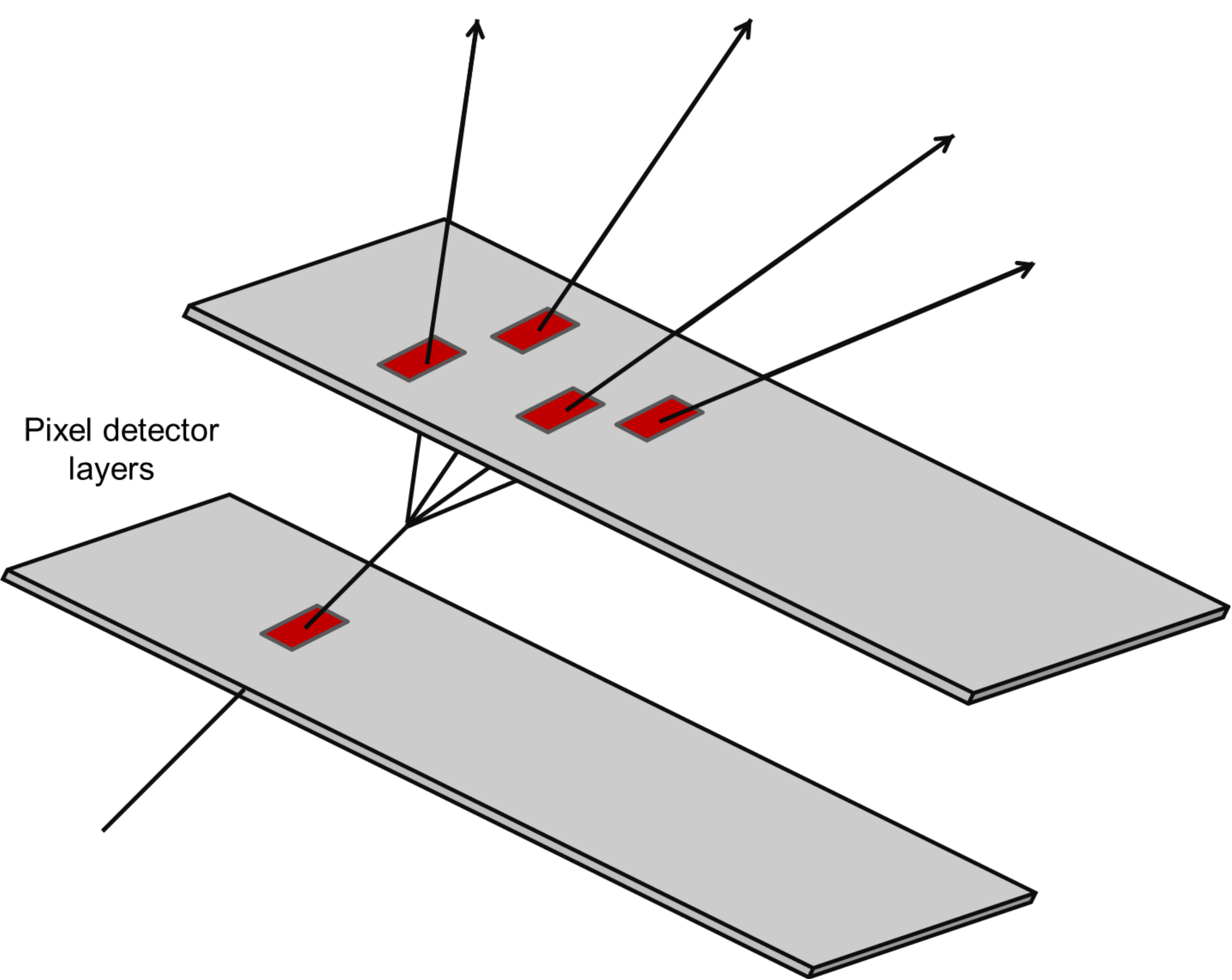}
\caption{The ``multiplicity jump'' tagger works when a particle
with a large Lorentz boost decays between two pixel layers. 
Shown here schematically is a particle traversing a pixel layer
from the lower left and decaying before the next layer, causing multiple
hits to appear. If highly boosted, $B$ hadrons have this property.}
\label{fig:BotOxJump}
\end{figure}

The principle underlying our previous study (see\cite{Huffman:2016wjk} 
and references therein) is the 
fact that in hadron colliders, a $B$ hadron with $\gamma \beta >200$ and 
typical proper lifetime in the $\approx 1$~ps range, will 
have a significant probability of passing through the inner layers of 
the detectors that surround the interaction point 
(see Figure~\ref{fig:BotOxJump}). Since the assumption 
underlying many tracking algorithms is that any fiducial 
track ought to leave a signal all the way to 
the inner layers, there is a significant chance that the tracks found would 
be missing hits on the inner layers, or might even have hits assigned 
incorrectly. 

A basic {\sc Geant4} detector simulation was used consisting of four cylindrical layers
of silicon which were segmented to approximate a typical pixel detector as
realized in LHC experiments such as ATLAS or CMS. {\sc Pythia8+EvtGen} were used to 
simulate the physics process where TeV mass $Z^\prime$ particles were created in order to 
produce high energy jets with $b$, $c$, and light quark progenitors. {\sc FastJet} was 
used to identify the jet axis using the anti-$k_T$ algorithm with a cone 
$R=0.2$.\footnote{A right-hand cylindrical coordinate system is adopted in this note 
where the $z$ axis points along the beam line, $r$ and $\phi$ are the radius and 
azimuthal angle in the plane transverse to the 
beam. $\eta \equiv -\ln{\tan{(\theta/2)}}$}

A ``hit'' in the detector was defined 
as a pixel segment in which at least 0.05 MeV had been deposited.  
Hits within a narrow cone around the jet axis of 
$\Delta R\equiv\sqrt{\Delta \eta^2+\Delta\phi^2}<0.04$
were used to define the relative multiplicity jump $f_j$ between layers $j$ and $j+1$, 
\begin{equation}
f_j=\frac{N_{j+1}-N_j}{N_j}=\frac{\Delta N_j}{N_j},
\label{eq:reljump}
\end{equation}
where $N_j$ is the number of hits in pixel layer $j$.
A jet was tagged as a $b$ jet if $f_j$ equaled or exceeded
a value $1$ for any layer $j$.  
The resulting tagging efficiency as a function of the parent jet energy showed 
clear separation between the tagging of $b$ over light-quark jets, and that 
the $b$-tagging efficiency remained fairly stable up to roughly 1.5~TeV.
This promising result, however, was obtained in
simulations without additional pile-up interactions or a finite luminous region 
along the beam line.  It was also suggested that it would be
worthwhile to investigate if machine learning techniques could further optimise the 
combination of the new $f_j$ observables. This paper addresses these points.
%\begin{figure}[thbp]
%\centering 
%%\vspace{-0.1 cm}
%\includegraphics[height=.65\textwidth]{R020ExtraMatEfficVsJetEnergyCut1p0.pdf}
%\caption{\label{fig:HitvsE} Efficiency of multiplicity jump tagging of fiducial
%$b$ jets as a function of jet energy, using all pixel layers and
%$\max(f_j)\geq 1$.  The dashed line indicates 600 GeV, the ``extreme energy''
%beyond which the efficiency of traditional $b$-tagging falls.  Also shown
%is the percentage of light-quark jets mis-tagged with the same cut. Note that 
%``efficiency'' as defined in this plot is only the efficiency of finding a $b$ jet where
%the $B$ hadron decayed within the fiducial volume of the pixel detector. }
%\end{figure}

% May also be a better way to optimise on the different pixel sizes in different layers.

\section{Simulation}
\label{sec:improv}

The authors continued to use a simulation based on {\sc Geant4} (version 10.0)
to model particle interactions and showering in a 
detector\cite{Agostinelli:2002hh}\cite{Allison:2006ve}.  
{\sc Pythia} version 8.209\cite{Pythia8}, with the default
Monash 2013 tune\cite{Skands:2014pea}, was used to simulate
$pp$ collisions with center-of-mass energy $\sqrt{s}=13\;{\rm
TeV}$.
Hard QCD events were generated with a Poisson mean of 45 soft QCD (minimum bias)
pile-up interactions for each hard QCD collision. 
The hard QCD process was set to have a minimum
$P_T > 700$~GeV for the underlying tree-level interactions,
and reconstructed jets with $P_T>350$~GeV from {\sc FastJet} were used.

Hard QCD from {\sc Pythia} was also used to create a pure sample 
of 300,000 $b$ jets
(also with pile-up) that were used to enrich the $B$ hadron content of the sample.
Simulations indicate the $B$ hadron takes most of the jet energy, with
the most likely energy fraction being $\simeq 85\%$, independent of the 
primary $b$ parton energy\cite{Peterson:1982ak}.
Decays of $B$ hadrons were simulated using {\sc EvtGen} version 1.4.0, with
bremsstrahlung handled by {\sc Photos} version 3.52 and any $\tau$ decays by
{\sc Tauola} version 1.0.7\cite{Lange:2001uf}. All of the events (hard QCD and 
pile-up) had collisions distributed along the beam-line with a Gaussian probability 
distribution having a width of $45$~mm. 

A simplified detector geometry, loosely based on the four-layer ATLAS pixel barrel
system, was used to model the detector response. The active pixel layers, with radii
25.7, 50.5, 88.5, and 122.5~mm, were encased within a volume of air and inside
a uniform 2~T magnetic field pointing in the positive $z$
direction. Each barrel was 1.3~m long (the innermost layer, the ``Insertable $B$
Layer'' or IBL, of the ATLAS pixel system is actually slightly
shorter\cite{AtlasTDR}). The pixel sensors were $300\;{\rm \mu m}$ 
thick, with a $50\;{\rm \mu m}$ pitch in the
$\phi$ direction, and a $400\;{\rm \mu m}$ length in the $z$
direction ($250\;{\rm \mu m}$ in our innermost layer, similar to the IBL in ATLAS). 
These idealized pixels were simulated as
pure silicon slabs without gaps. 

In order to model inactive material, further cylinders of
silicon were added to the {\sc Geant4} model, located just outside each 
cylinder of sensitive pixels, so as to
bring the total simulated material up to an equivalent of 2.5\%
of radiation length per layer. In addition a silicon cylinder half as thick
was added just inside the
outermost active layer of pixels.

Stable generated particles (excluding neutrinos) were clustered using the 
{\sc FastJet} (version 3.1.3)\cite{FastJet} implementation of the ``anti-$k_T$''
sequential recombination algorithm\cite{antikt} with $R=0.2$.
The jet's axis was corrected for the position of the primary vertex and 
was then used to define the angular search region in the multiplicity jump algorithm.

The sample of $b$ jets was defined by finding the highest energy ground state
$B$ hadron within $\Delta R<0.2$ of the jet axis. After $b$ jets
were identified, a similar search was performed to identify charm jets. All other
jets were considered
``light quark'' jets (or ``uds'') jets. The two
highest energy $b$ jets were then used to test the efficiency of
the multiplicity jump algorithm. 

\section{Performance}
 \label{sec:Perform}

\subsection{Effect of pile-up}

\begin{figure}[thbp]
\centering 
%\vspace{-0.1 cm}
\includegraphics[height=.75\textwidth]{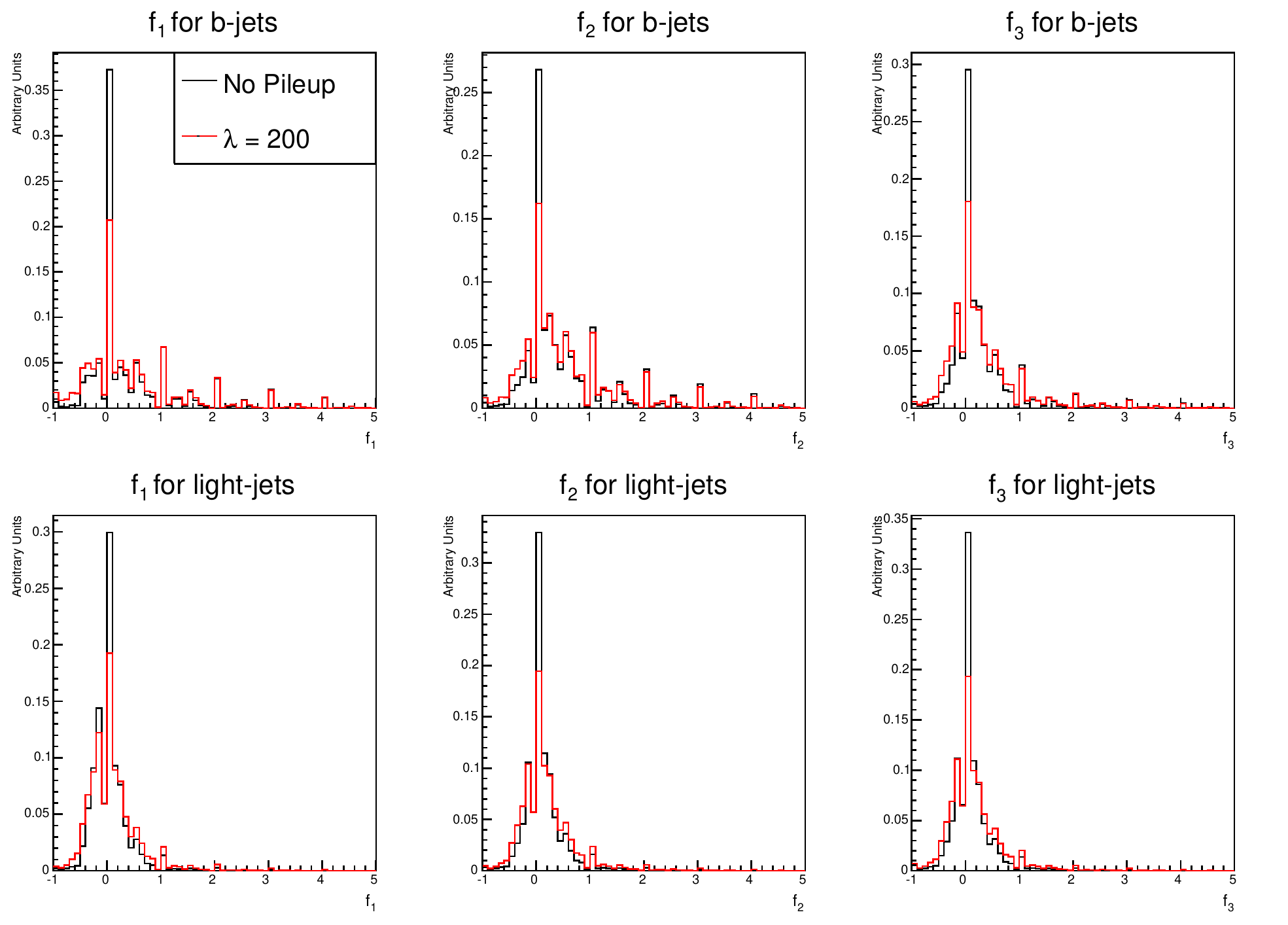}
\caption{\label{fig:pile-up} Shown are the histograms of the quantity $f_j$ for each of the 
three gaps within the simulated pixel detector comparing the case with no pile-up (black)
and with an average of $\lambda=200$ pile-up events (red). The probability of 
obtaining a negative or zero value for $f_j$ has increased; however the key feature, that $b$ jets
have a substantial positive tail compared to uds-jets, remains.}
\end{figure}
Figure~\ref{fig:pile-up} shows histograms of $f_j$ from equation~\ref{eq:reljump} for 
$b$ and light-quark jets for no pile-up
and events with an average of $\lambda=200$ (rather than the nominal 45) additional interactions
spread over the luminous region.  It is clear that, having corrected the jet axis
for the primary vertex position, the narrow search cone, $\Delta R<0.04$, excludes most
additional pile-up hits.  The histograms indicate that the main effect of pile-up 
is to reduce the chance that $f_j = 0$, which only occurs when the number of hits
between the layers in question within the search cone are the same. Comparing 
the $b$ jet and uds-jet cases one can see that, regardless of pile-up, the $b$ jets still
have long tails for $f_j \geq 1.0$. 
Since this region of $f_j$ is largely
insensitive to pile-up, we conclude that its effects on cut and neural net-based approaches
to multiplicity jump tagging are similarly insensitive.  Nevertheless, we
continued to simulate an average of 45 pile-up interactios per event for
other studies in this article.

\subsection{Artificial Neural Network}
\label{sec:ANN}

Both CMS and ATLAS have already shown the benefits of using 
multivariate techniques such as boosted decision trees and artificial neural 
networks (ANN's) for $b$-tagging at the LHC\cite{AtlasMV2}.\footnote{An ANN is also referred to as a 
``multilayer perceptron'' in some of the literature.}
It is a natural next step to see what improvements 
might be possible within the multiplicity jump technique.

\begin{figure}[thbp]
\centering 
%\vspace{-0.1 cm}
\includegraphics[height=.55\textwidth]{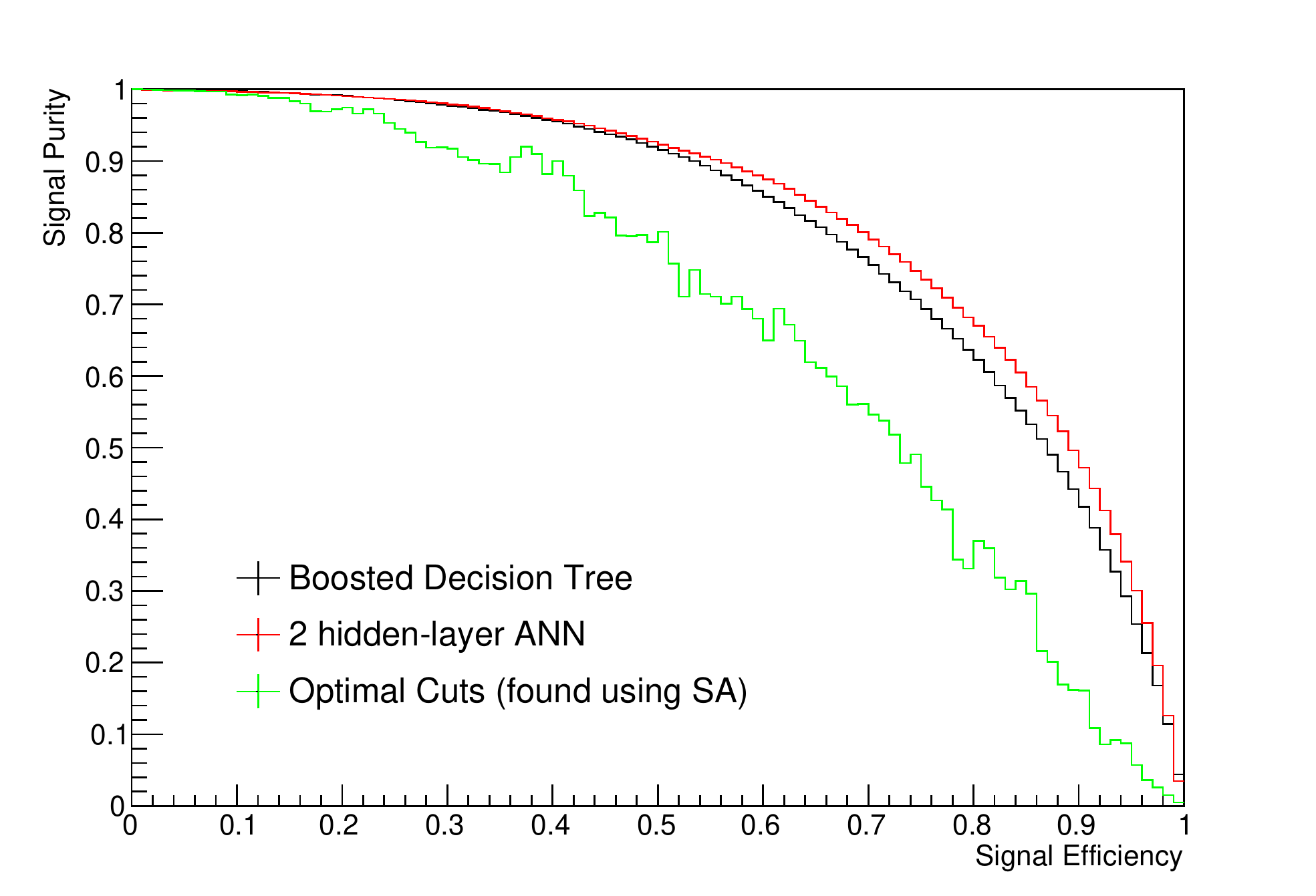}
\caption{\label{fig:ROC} A comparison of the efficiency vs. purity of the ANN,
a boosted decision tree, and the simple cut-based technique 
from reference\cite{Huffman:2016wjk}.}
\end{figure}
The Toolkit for Multivariate Analysis (TMVA) packaged 
within ROOT (version 5.34.21)\cite{Hocker:2007ht} 
was used to evaluate a set of multivariate analysis techniques in parallel, giving
one information about the efficiency at various purity levels, and the integral
of the efficiency-purity (the ``ROC'') curve.  
Figure~\ref{fig:ROC} shows a comparison of ROC
curves for a Boosted Decision Tree (BDT), an ANN with two hidden layers (described below), and a
cut-based analysis. Simulated annealing (SA) is used to determine the optimal set of cut values 
for a given efficiency. The ANN approach is preferred in this evaluation.

In figure~\ref{fig:ROC}, the BDT depth was 200 nodes. 
So the usual advantages (such as the ability to deduce its inner workings) typically conferred
by decision trees over neural networks did not apply.  Moreover, while
a BDT is often faster to execute once trained, an ANN with only two internal layers
is shallow enough that the difference in execution time is reduced, and could be
further reduced using standard parallelism techniques.

The main ANN chosen has a single input layer with $N$ neurons, where $N$ is the number of 
input variables. There follow two hidden layers, the first with $N+5$~neurons and 
the second with 5~neurons. 
This second layer is connected to a single output neuron whose output
spans the interval [0,1] and is used as the discriminant. The activation function used in the 
hidden layers is $f(x) = \tanh{(x)}$, where $x$ is the weighted sum of the outputs of the 
previous layer.
Tests using up to eight hidden layers 
responded with identical results on identical samples as two layers. 
At the same time, two layers were chosen in order to satisfy the requirements of the 
universal approximation theorem\cite{Hornik:1991NN}.

The list of inputs is as follows:
\begin{itemize}
\item jet $P_T$, energy, and mass from the anti-$k_T$ jet; \\
\item The raw hit numbers in each layer within the search cone of $R<0.04$ of the jet axis;\\
\item each of the $f_i$ as defined in equation~\ref{eq:reljump}; and \\
\item the maximum of the $f_i$ (max($f_i$)) over all the layers.\\
\end{itemize}

Note the input variables overlap somewhat: 
in principle an ANN should be able to use the $f_i$ automatically to obtain the 
equivalent discrimination power that max($f_i$) provides. 
However, prior knowledge applied to 
the inputs to an ANN will improve the efficiency of the training phase: using 
max($f_i$) as an input saves the ANN from having to derive an equivalent. 
At the same time, should max($f_i$) prove of little use, the ANN should  
de-weight that input in favour of more powerful combinations. 

The neural network is trained using standard back-propagation techniques on a dataset 
of 1~million hard QCD events (including $b$ jets) 
and enriched with 300,000~hard QCD events that resulted 
in $B$ hadron production. For training, 
a ``signal'' event was a $b$ jet with a $B$ hadron that decayed 
within the fiducial volume. A ``background'' event was anything that was not a $b$ jet.

\begin{figure}[thbp]
\centering 
%\vspace{-0.1 cm}
\includegraphics[height=.55\textwidth]{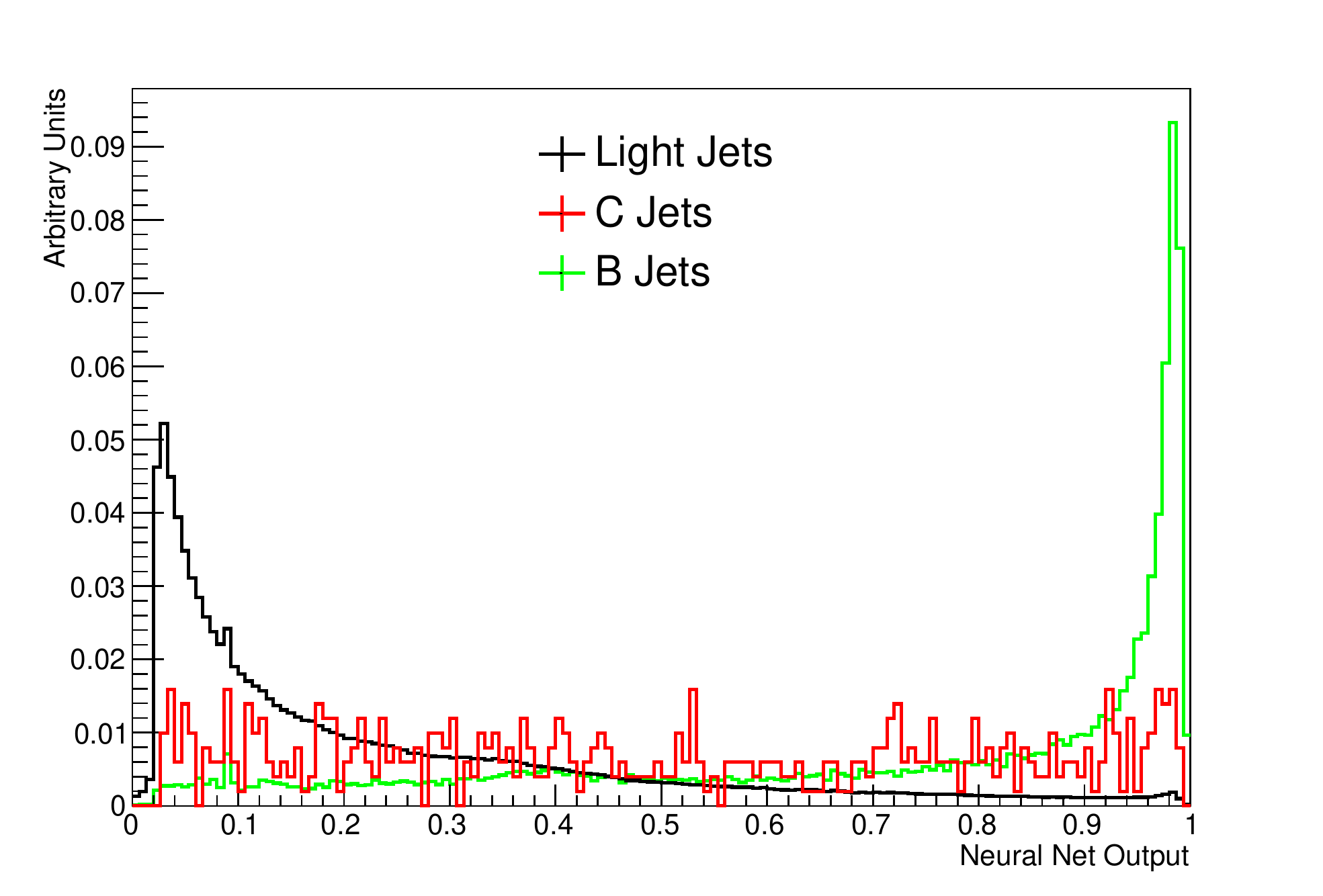}
\caption{\label{fig:NNoutput} The distribution of the ANN output. 
$b$, $charm$, and $uds$ jets are identified by colour. The charm and bottom jets plotted here
are only those jets where the final state hadron decayed within the fiducial volume.}
\end{figure}

The test sample consisted of 1 million hard QCD events with an average of 45 pile-up events, 
generated in the same way but with a different random number
seed. No additional $b$ jets were added. 
The resulting 
discriminant distribution is shown in figure~\ref{fig:NNoutput}.
$b$ jets are strongly peaked towards a value of 1, 
while uds jets are strongly peaked towards low values. 
Charm jets which decay within the fiducial region, on the other hand, are distributed 
roughly evenly across the discriminant range, which is not surprising since the ANN was 
trained to recognize only $b$ jets as signal.

\subsection{ANN results}

In order to measure the algorithm's performance in our simulation, we define
an efficiency $\epsilon_b$ for $b$ jets in a fiducial region as
the number of tagged $b$ jets divided by the number of jets in which the
matched $B$ hadron decays within the fiducial region. The fiducial region is
defined in terms of the inner and outer pixel layers being investigated;
in other words, the $\epsilon_b$ reflects the probability that,
if a $B$ hadron decays between the inner and outer layers, it will be 
tagged by the algorithm.
The light-quark ``efficiency'' $\epsilon_q$ is the number of
light-quark jets tagged by the algorithm divided by the number of
light-quark jets. 

To see how the tagging efficiency behaves as a function of the jet energy 
for a given cut on the ANN output, we use a figure of merit designed to optimize the 
significance of the tagged B jets over the light quark jets. (Charm jets are explicitly 
excluded at this point.) The figure of merit used is 
\begin{equation}
S = \frac{\epsilon_b}{\sqrt{\epsilon_q }}
\label{eq:sig2}
\end{equation}
where $\epsilon_b$ and $\epsilon_q$ are the efficiency of the tag on $b$ jets 
and the mistag rate for light quark jets, respectively. There are a few different ways
of quantifying significance-like values. In this case 
$S$ has the advantage of being independent of the absolute numbers of the 
different types of jets and should be independent of specific underlying 
jet production mechanisms. 

\begin{figure}[thbp]
\centering 
%\vspace{-0.1 cm}
\includegraphics[height=.55\textwidth]{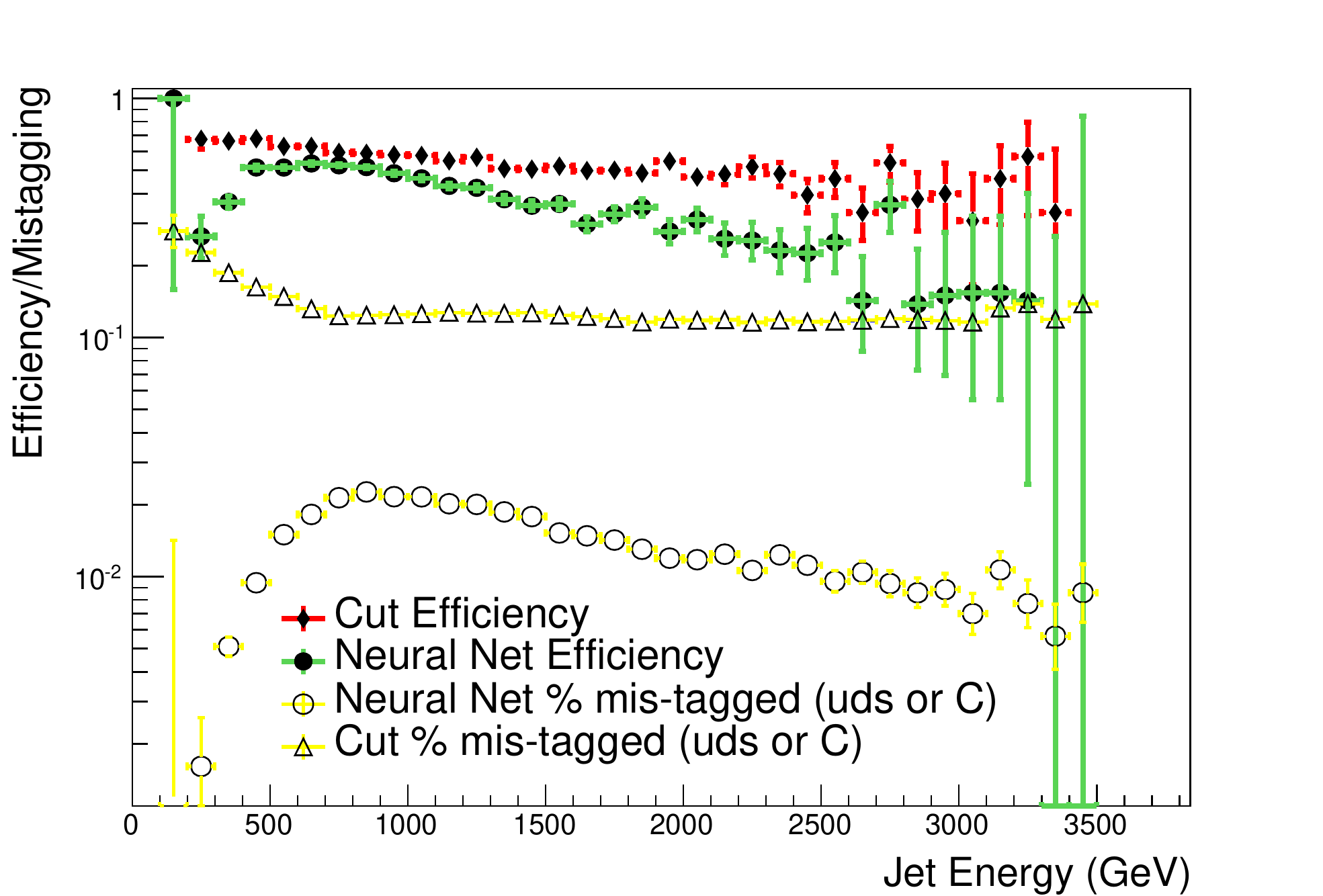}
\caption{\label{fig:effresult} Efficiencies and mistag rates for the ANN 
and cut-based tagging algorithms. The ANN algorithm uses a value greater 
than $0.9$ from figure~\ref{fig:NNoutput}, while
the cut-based algorithm tags jets with $\max(f_j) \geq 1$.}
\end{figure}

%\begin{figure}[thbp]
%\centering 
%\includegraphics[height=.5\textwidth]{neural-net-significance-plot.pdf}
%\caption{\label{fig:ANNsig} Shown here is the ``significance'' as defined in 
%equation~\ref{eq:sig2}. Events with an ANN output $>0.9$ were used 
%to make the efficiency plot in figure~\ref{fig:effresult}.}
%\end{figure}
%One can choose the point of operation for the ANN from the ROC curve in 
%Figure~\ref{fig:ROC} by maximizing the figure of merit $S$ in 
%equation~\ref{eq:sig2}. 
Figure~\ref{fig:effresult} shows the $b$ jet efficiency and the mistag 
percentage of light-quark
jets vs. jet energy for an ANN output value greater than $0.9$. 
The cut-based results on the same sample are also shown. Choosing an 
ANN output $>0.9$, where the significance defined in equation~\ref{eq:sig2} 
is near the maximum, optimizes the statistical power of finding $b$ jets over
light quark jets. With this optimization the loss in efficiency is compensated by 
the ability to discriminate against uds jets. 

\begin{figure}[thbp]
\centering 
\includegraphics[height=.55\textwidth]{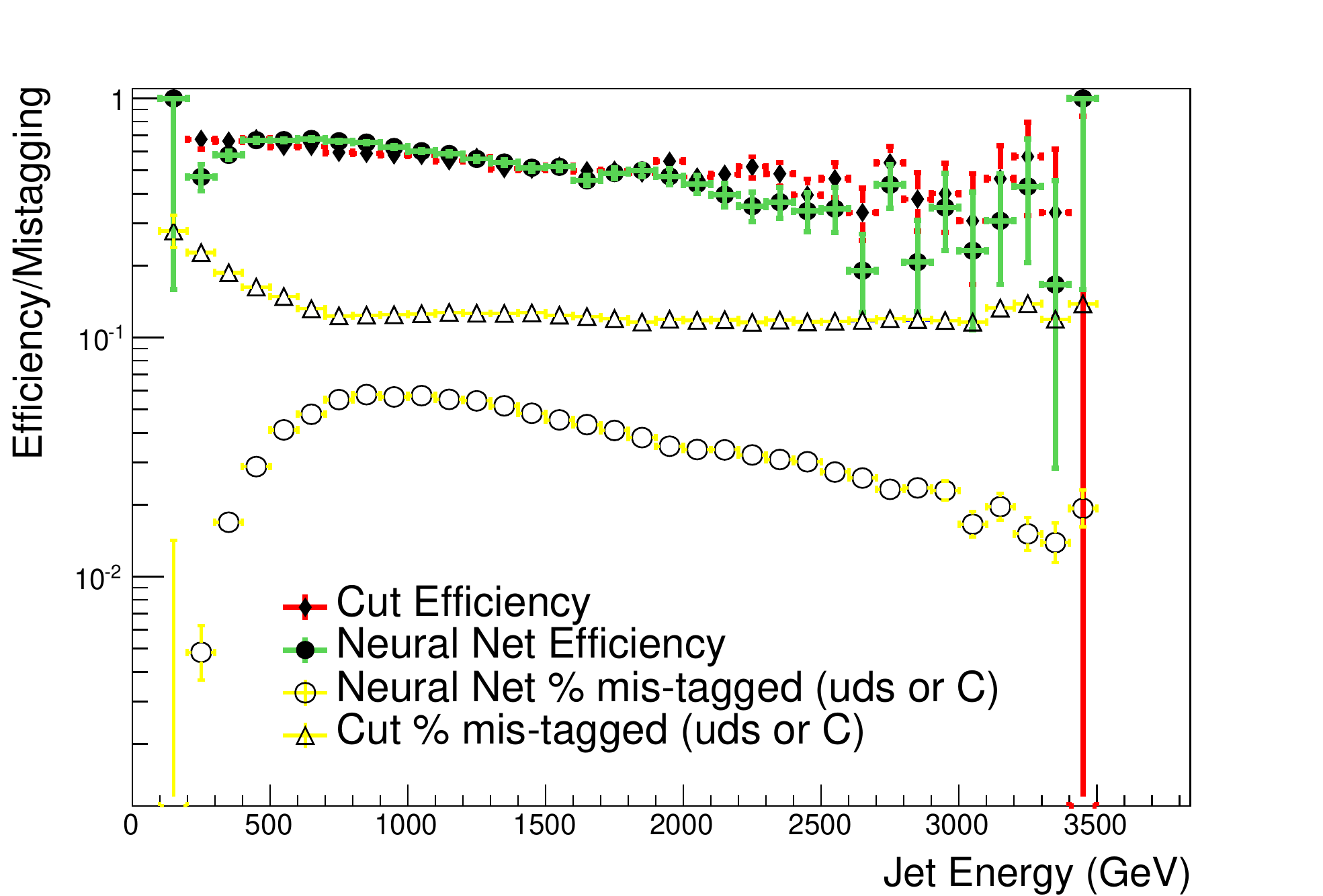}
\caption{\label{fig:effmatch} Efficiencies and mistag rates for the ANN and cut-based tagging
algorithms as in figure~\ref{fig:effresult}. Here a value of the ANN output greater than 0.75
is set so as to match the cut-based b-tagging efficiency.}
\end{figure}

When the highest significance is used, the absolute efficiency obtained from the ANN 
does not exceed what was achieved with a cut-based approach using a similar 
significance metric. 
So we also show a case with the ANN discriminate greater than 0.75, 
chosen to match the $b$ jet efficiencies of the ANN and 
cut-based algorithms. This is shown in figure~\ref{fig:effmatch} where the 
multivariate technique still outperforms the cut-based method in rejecting uds jets.

\section{Cross checks and Further tests}
\label{sec:further}

Artificial neural networks are difficult to characterize and can lead to false
training, where the quantity the user thinks is providing the discriminating power
is actually being bypassed in favor of a nonsensical combination.  One concern here 
is that, because the $B$ hadron fraction might be a function
of the jet energy, the ANN could rely too heavily on
this or other energy-related quantities.  One of the reasons 
high mass $Z^{'}$ states were chosen 
for both the signal and background jet samples is so that the energy spectrum of the 
resulting jets would be the same. 

In order to test for this possibility,
the training was repeated on a hard QCD jet sample with a flat energy distribution
from 1 to 6 TeV.  After re-training, however, the efficiency and mistag rates for
a given cut on the discriminant were virtually the same and behaved similarly
as a function of jet energy, suggesting that false training on jet energy,
at least, was not a problem.
Additionally tests we performed upon the trained ANN show that changes in the 
jet energy, jet mass, and jet $P_T$ have at least an order of magnitude smaller impact 
on the ANN output compared to inputs like $f_i$ or max($f_i$) which are related to hits.
(See subsection~\ref{subsec:heatmap} below.) 

The effect of the inactive material mentioned in section~\ref{sec:improv} 
on mistag rates was investigated by removing the
inactive material and generating a new test sample.  The ANN was not
re-trained.  The mistag rates for both cut-based and ANN-based analyses
fall, but that for the ANN falls much more. The overall $b$-tagging efficiency also
drops somewhat, possibly due to the fact that the ANN was trained on a
sample simulated with the extra material.

The ANN-based tagger, like the cut-based tagger, exhibits a fall in
efficiency at jet energies above 1 TeV.  Since the efficiency as defined
above counts only those $B$ hadrons which could in principle have been
detected by a multiplicity jump, any loss of $B$ hadrons because they live too
long ought to have been removed.  However, since some 85\% of $B$ hadrons
decay into charm hadrons with a similar lifetime, it is suspected that the
ANN uses combinations of hits in different layers that would benefit from
the charm decay; with highly boosted hadrons, it is more likely that
the charm daughter would live long enough to escape the pixel volume.
This hypothesis was tested by adding a fifth pixel layer at higher radius, and it was
observed that the $b$-tagging efficiency was maintained for higher
energy jets.

\subsection{Input weights} \label{subsec:heatmap}
\begin{figure}
  \centering
  \includegraphics[scale=1.25]{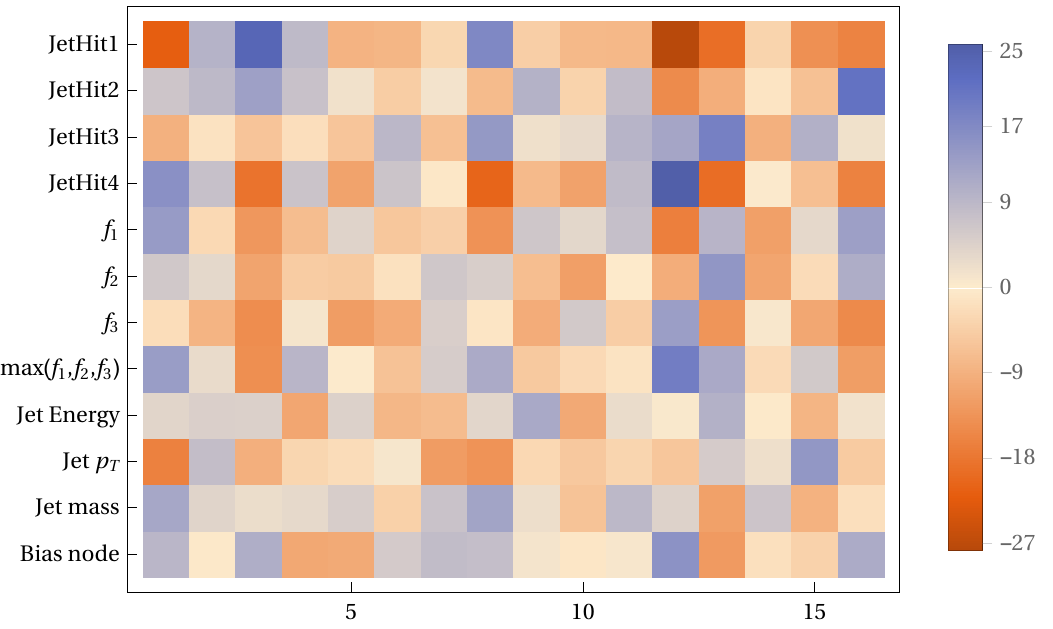}
  \caption{A ``heatmap'' of the weights $w_{ij}$ between the input and first hidden layers. The 
	colors are an indication of how strongly positively or negatively the given input is influencing 
	a given internal node of the ANN. Note the near complete anti-correlation between the number 
	of hits in layer 1 ({\it JetHit1}) and layer 4 ({\it JetHit4}) for many internal nodes.}
  \label{subfig:heatmap}
\end{figure}

To gain insight into which inputs were being most heavily weighted after training, we
 generated maps of the weights between the input and the ANN's first hidden layer. 
This is shown in Figure~\ref{subfig:heatmap}, where the weight that the trained 
neural net assigns to an input has been indicated by a color, with a darker color indicating 
a high weight.
By looking at Figure~\ref{subfig:heatmap} we can see that all of 
the kinematic variables (like jet energy) have small weights. This indicates, in addition to the 
studies explained previously, that the jet-related inputs have small individual effects.
	
Figure~\ref{subfig:heatmap} shows that the $JetHit1$ and $JetHit4$ 
variables were heavily weighted. Moreover they are anti-correlated, i.e. when $JetHit1$ 
is heavily positively weighted, $JetHit4$ is highly negatively weighted, and vice versa.
	
\begin{figure}
  \centering
  \includegraphics[scale=0.65,angle=0]{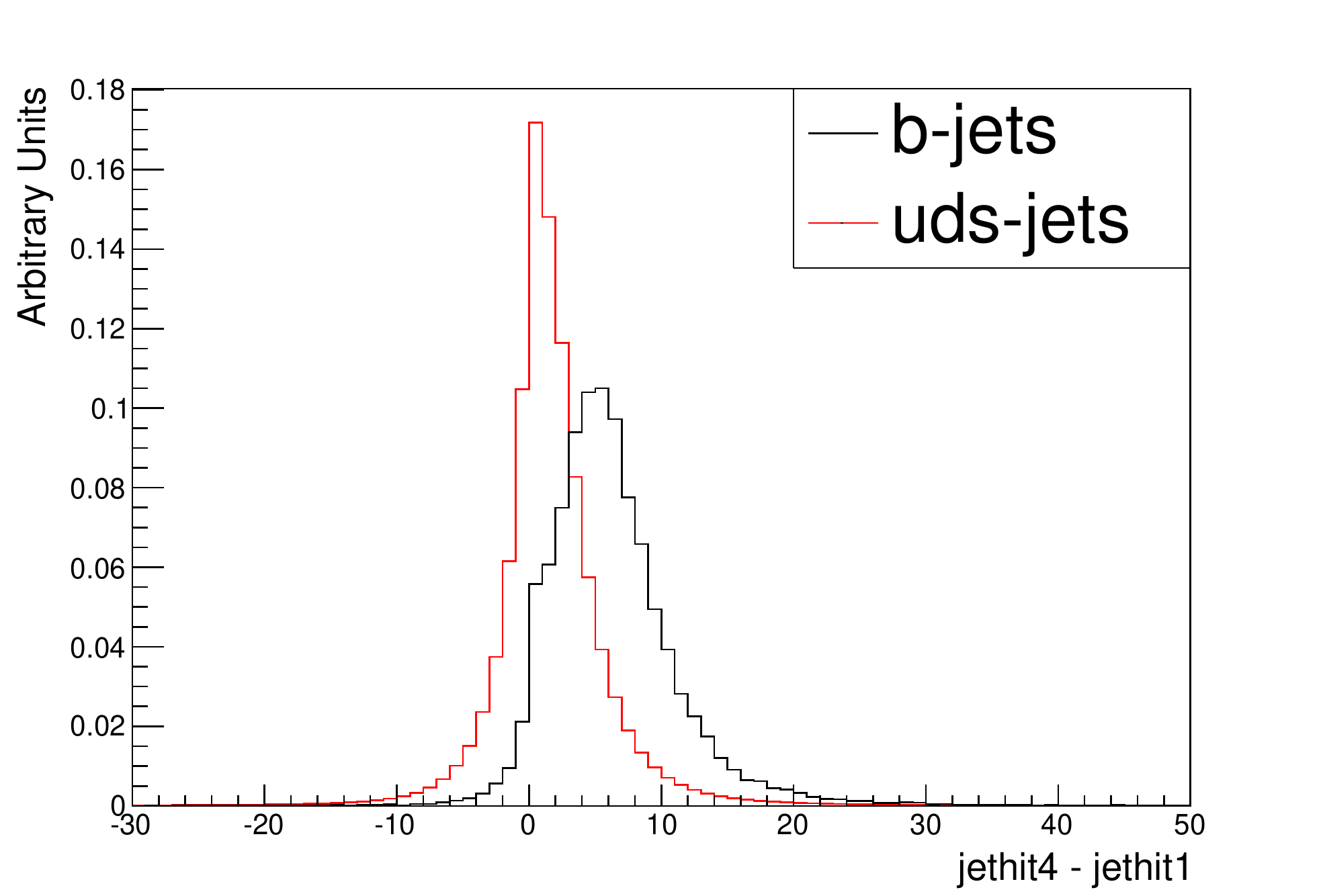}
  \caption{Plot showing $|JetHit4 - JetHit1|$ for $b$ and uds jets. Based upon this plot and the 
	heatmap of figure~\ref{subfig:heatmap} it appears the ANN found a similar relationship 
	to aid in tagging $b$ jets.}
  \label{fig:jethit4-jethit1}
\end{figure}

This analysis invites us to reason that some combination of 
$|JetHit4 - JetHit1|$ is being used as a primary discriminant within the neural network. 
To investigate this, we show this hit difference for $b$ jets and uds jets in 
Figure~\ref{fig:jethit4-jethit1}. This showed that there was a significant separation in 
the distributions for $b$ and uds jets and so implied that it could be a powerful individual tagger.
	
Finally, in order to fully satisfy ourselves that the jet kinematic variables were not having undue 
individual effect on the ANN, we calculated the analytic function for the ANN output in Mathematica. 
The partial derivatives of the ANN output with respect to the various inputs were computed.
As expected, in general the derivatives for a given input are large where the heat map 
indicates a large effect on the output.

\subsection{Changing the search cone}
As mentioned in section~\ref{sec:rehash} we used a fixed search cone for pixel hits of 
$\Delta R\leq0.04$
in order to make comparisons to~our previous work transparent\cite{Huffman:2016wjk}. 
But it is also clear from the previous studies that lower energy $b$ jets contain 
lower energy $B$ hadrons.  In the Monte Carlo samples that we used $\Delta R\leq0.04$
only contained about 85\% of the $B$ hadrons. 
Indeed, the best solution would be to allow the ANN to choose hits within a variable cone 
size that would depend on the energy of the jet. 

\begin{figure}
  \centering
  \includegraphics[scale=0.9,angle=0]{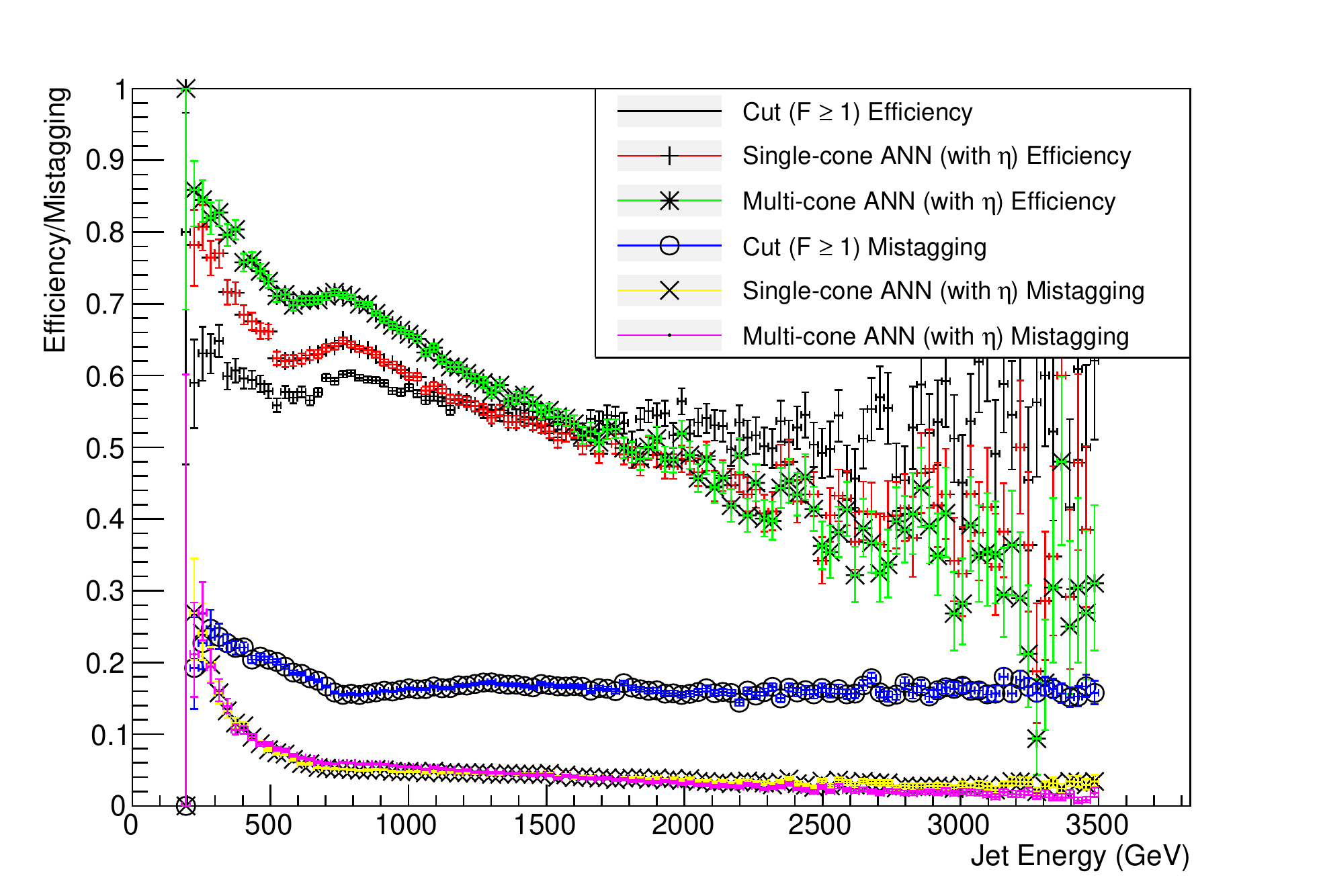}
  \caption{Shown is a comparison of the efficiency of the fixed-cone ANN from 
	figure~\ref{fig:effresult} and the ANN where mulitple inputs with search cones for hits of 
	$R\leq0.10$ and $R\leq0.20$ were used in addition to the original search cone of $R\leq0.04$.
	The 2 and 4-layer ANN outputs were chosen such that they matched in light quark rejection 
	making a comparison of the $b$ jet tagging effectiveness more obvious.}
  \label{subfig:multiconeEff}
\end{figure}

To do something very similar to an energy-dependent cone search 
new input discriminators, $JetHitN$ and $f_N$ (where $N$ is the layer number), 
were added to a new artificial neural net, each corresponding to cone sizes
of $\Delta R\leq0.04$, $\Delta R\leq0.11$, and $\Delta R\leq0.20$ thus increasing the 
number of inputs.  
Repeating the previous study for the number of internal layers indicated that a 4-layer ANN was
needed to accomodate the increased complexity of inputs. 
This modified ANN was implemented and trained using a much larger 
training sample.
Figure~\ref{subfig:multiconeEff}
shows the case where the 4-layer ANN output is chosen to match the light-quark rejection 
of the two-layer ANN case. One can see that improvements occur in the lower energy jets, consistent 
with the use of larger cone sizes. As the jet energy increases the benefits of having larger cones 
lessens such that there is no difference between the cases at the highest jet energies. 

The training requirements for the multi-cone ANN were substantially more intensive than the 
previous 2-layer case. Training for the ANN that generated figure~\ref{fig:effresult} required several 
hours while the training for the 4-layer case took 3 days on the same computer system.  Training 
requirements could be a limitation for more complex neural nets. 

\section{Conclusions and further study}
\label{sec:conclusion}

This article focuses on using an artificial neural network (ANN) to
optimize the discriminating power that may be afforded by a jump in
the numbers of hits from the inner to outer layers of a pixel detector
in the presence of a highly boosted $B$ hadron.

This study accounts for pile-up at a similar level, and spread over a broad
luminous region, as currently experienced by the LHC experiments.
Additional studies using more internal node layers to accomodate a 
range of input variables that account for differing cone sizes showed 
improvements to $b$-tagging over a fixed cone size, but with the 
disadvantage of increased training requirements. 

Other complications arising from realistic detector geometry,
including overlaps between sensors comprising the same layer, and the
transition from cylindrical to endcap disk layers, have still to be
investigated.  Nonetheless, the use of a multivariate technique has
shown a significant improvement in the mistag rate for jets that did
not contain a $B$ hadron.  Training on charm jets with the addition
of another output neuron may recover charm tags as well.

As noted in \cite{Huffman:2016wjk}, if shown to work in the LHC detectors 
this technique could have
implications for the detector
design at future colliders such as the Future Circular Collider
(FCC)\cite{FCC}. Extending finely segmented pixel coverage to larger radii 
in order to tag these jets may be desirable for such future
detectors.

\section*{Acknowledgments}

The authors would like to thank Cigdem Issever for her encouragement to publish 
these findings. 
This work was supported by the Science and Technology Facilities
Council of the United Kingdom grant number ST/N000447/1 
and the Higher Education Funding Council of England.

\section*{References}

\bibliography{JumpMultbibTCJ2p2}{}

\providecommand{\href}[2]{#2}\begingroup\raggedright\begin{thebibliography}{10}

\bibitem{Bishara:2016kjn}
F.~Bishara, R.~Contino and J.~Rojo, \emph{{Higgs pair production in
  vector-boson fusion at the LHC and beyond}},
  \href{http://arxiv.org/abs/1611.03860}{{\tt 1611.03860}}.

\bibitem{Behr:2015oqq}
J.~K. Behr, D.~Bortoletto, J.~A. Frost, N.~P. Hartland, C.~Issever and J.~Rojo,
  \emph{{Boosting Higgs pair production in the $b\bar{b}b\bar{b}$ final state
  with multivariate techniques}},
  \href{http://dx.doi.org/10.1140/epjc/s10052-016-4215-5}{\emph{Eur. Phys. J.}
  {\bf C76} (2016) 386}, [\href{http://arxiv.org/abs/1512.08928}{{\tt
  1512.08928}}].

\bibitem{Gouzevitch:2013qca}
M.~Gouzevitch, A.~Oliveira, J.~Rojo, R.~Rosenfeld, G.~P. Salam and V.~Sanz,
  \emph{{Scale-invariant resonance tagging in multijet events and new physics
  in Higgs pair production}},
  \href{http://dx.doi.org/10.1007/JHEP07(2013)148}{\emph{JHEP} {\bf 07} (2013)
  148}, [\href{http://arxiv.org/abs/1303.6636}{{\tt 1303.6636}}].

\bibitem{Alwall:2008ag}
J.~Alwall, P.~Schuster and N.~Toro, \emph{{Simplified Models for a First
  Characterization of New Physics at the LHC}},
  \href{http://dx.doi.org/10.1103/PhysRevD.79.075020}{\emph{Phys. Rev.} {\bf
  D79} (2009) 075020}, [\href{http://arxiv.org/abs/0810.3921}{{\tt
  0810.3921}}].

\bibitem{CMSbtag}
{\scshape CMS} collaboration, \emph{{Identification of b quark jets at the CMS
  Experiment in the LHC Run 2}},  Tech. Rep. CMS PAS BTV-15-001, CERN, Geneva,
  May, 2016.

\bibitem{ATLASbtag}
{\scshape ATLAS} collaboration, \emph{{Identification of Jets Containing
  $b$-Hadrons with Recurrent Neural Networks at the ATLAS experiment}},  Tech.
  Rep. ATL-PHYS-PUB-2017-003, CERN, Geneva, March, 2017.

\bibitem{ATLASbtag2}
{\scshape ATLAS} collaboration, \emph{{Optimisation of the ATLAS $b$-tagging
  performance for the 2016 LHC Run}},  Tech. Rep. ATL-PHYS-PUB-2016-012, CERN,
  Geneva, June, 2016.

\bibitem{Huffman:2016wjk}
B.~T. Huffman, C.~Jackson and J.~Tseng, \emph{{Tagging $b$ quarks at extreme
  energies without tracks}},
  \href{http://dx.doi.org/10.1088/0954-3899/43/8/085001}{\emph{J. Phys.} {\bf
  G43} (2016) 085001}, [\href{http://arxiv.org/abs/1604.05036}{{\tt
  1604.05036}}].

\bibitem{Agostinelli:2002hh}
{\scshape GEANT4} collaboration, S.~Agostinelli et~al., \emph{{GEANT4: A
  Simulation toolkit}},
  \href{http://dx.doi.org/10.1016/S0168-9002(03)01368-8}{\emph{Nucl. Instrum.
  Meth.} {\bf A506} (2003) 250--303}.

\bibitem{Allison:2006ve}
J.~Allison et~al., \emph{{Geant4 developments and applications}},
  \href{http://dx.doi.org/10.1109/TNS.2006.869826}{\emph{IEEE Trans. Nucl.
  Sci.} {\bf 53} (2006) 270}.

\bibitem{Pythia8}
T.~Sjostrand, S.~Mrenna and P.~Z. Skands, \emph{{A Brief Introduction to PYTHIA
  8.1}}, \href{http://dx.doi.org/10.1016/j.cpc.2008.01.036}{\emph{Comput. Phys.
  Commun.} {\bf 178} (2008) 852--867},
  [\href{http://arxiv.org/abs/0710.3820}{{\tt 0710.3820}}].

\bibitem{Skands:2014pea}
P.~Skands, S.~Carrazza and J.~Rojo, \emph{{Tuning PYTHIA 8.1: the Monash 2013
  Tune}}, \href{http://dx.doi.org/10.1140/epjc/s10052-014-3024-y}{\emph{Eur.
  Phys. J.} {\bf C74} (2014) 3024}, [\href{http://arxiv.org/abs/1404.5630}{{\tt
  1404.5630}}].

\bibitem{Peterson:1982ak}
C.~Peterson, D.~Schlatter, I.~Schmitt and P.~M. Zerwas, \emph{{Scaling
  Violations in Inclusive e+ e- Annihilation Spectra}},
  \href{http://dx.doi.org/10.1103/PhysRevD.27.105}{\emph{Phys. Rev.} {\bf D27}
  (1983) 105}.

\bibitem{Lange:2001uf}
D.~J. Lange, \emph{{The EvtGen particle decay simulation package}},
  \href{http://dx.doi.org/10.1016/S0168-9002(01)00089-4}{\emph{Nucl. Instrum.
  Meth.} {\bf A462} (2001) 152--155}.

\bibitem{AtlasTDR}
M.~Capeans, G.~Darbo, K.~Einsweiller, M.~Elsing, T.~Flick, M.~Garcia-Sciveres
  et~al., \emph{{ATLAS Insertable B-Layer Technical Design Report}},  Tech.
  Rep. CERN-LHCC-2010-013. ATLAS-TDR-19, CERN, Geneva, Sep, 2010.

\bibitem{FastJet}
M.~Cacciari, G.~P. Salam and G.~Soyez, \emph{{FastJet User Manual}},
  \href{http://dx.doi.org/10.1140/epjc/s10052-012-1896-2}{\emph{Eur. Phys. J.}
  {\bf C72} (2012) 1896}, [\href{http://arxiv.org/abs/1111.6097}{{\tt
  1111.6097}}].

\bibitem{antikt}
M.~Cacciari, G.~P. Salam and G.~Soyez, \emph{{The Anti-k(t) jet clustering
  algorithm}},
  \href{http://dx.doi.org/10.1088/1126-6708/2008/04/063}{\emph{JHEP} {\bf 04}
  (2008) 063}, [\href{http://arxiv.org/abs/0802.1189}{{\tt 0802.1189}}].

\bibitem{AtlasMV2}
{\scshape ATLAS} collaboration, \emph{{Optimisation of the ATLAS $b$-tagging
  performance for the 2016 LHC Run}},  Tech. Rep. ATLAS-PHYS-PUB-2016-012,
  CERN, Geneva, June, 2016.

\bibitem{Hocker:2007ht}
A.~Hocker et~al., \emph{{TMVA - Toolkit for Multivariate Data Analysis}},
  {\emph{PoS} {\bf ACAT} (2007) 040},
  [\href{http://arxiv.org/abs/physics/0703039}{{\tt physics/0703039}}].

\bibitem{Hornik:1991NN}
K.~Hornik, \emph{{Approximation capabilities of multilayer feedforward
  networks}}, {\emph{Neural Networks} {\bf 4} (1991) 251--257}.

\bibitem{FCC}
{\scshape TLEP Design Study Working Group} collaboration, M.~Bicer et~al.,
  \emph{{First Look at the Physics Case of TLEP}},
  \href{http://dx.doi.org/10.1007/JHEP01(2014)164}{\emph{JHEP} {\bf 01} (2014)
  164}, [\href{http://arxiv.org/abs/1308.6176}{{\tt 1308.6176}}].

\end{thebibliography}\endgroup
\bibliographystyle{JHEP}

\end{document}